\documentclass[aps,twocolumn,prl,groupedaddress,showpacs]{revtex4-1}

\usepackage{graphicx,float}
\usepackage[normalem]{ulem}
\usepackage{color}
\usepackage{amssymb}
\usepackage{bbm}

\usepackage{calligra}
\DeclareMathAlphabet{\mathcalligra}{T1}{calligra}{m}{n}
\DeclareFontShape{T1}{calligra}{m}{n}{<->s*[2.2]callig15}{}

\def\be{\begin{equation}}
\def\ee{\end{equation}}
\def\e#1{\label{#1}\end{equation}}
\def\bea{\begin{eqnarray}}
\def\eea{\end{eqnarray}}
\def\ea#1{\label{#1}\end{eqnarray}}

\def\bem#1{\begin{mathletters}\label{#1}}
\def\eml{\end{mathletters}}

\def\ket#1{{|#1\rangle}}

\def\4#1{{\boldsymbol{#1}}}
\def\8#1{{\widetilde{#1}}}
\def\bse{\begin{subequations}}
\def\ese{\end{subequations}}

\def\Rb87{$^{87}\text{Rb}$}
\def\0{\ket{0}}
\def\1{\ket{1}}

\begin{document}
\title{Indirect quantum sensors: Improving the sensitivity in characterizing very weakly coupled spins}
\author{Johannes N. Greiner}
\author{D. D. Bhaktavatsala Rao }
\author{Philipp Neumann}
\author{J\"{o}rg Wrachtrup}

\affiliation{3. Physikalisches Institut, Research Center SCOPE, University of Stuttgart, Pfaffenwaldring 57, 70569 Stuttgart, Germany
}%
\date{\today}
\begin{abstract}
We propose a scheme to increase the sensitivity and thus the detection volume of nanoscale single molecule magnetic resonance imaging. The proposal aims to surpass the $T_1$ limited detection of the sensor by taking advantage of a long-lived ancilla nuclear spin to which the sensor is coupled. We show how this nuclear spin takes over the role of the sensor spin, keeping the characteristic time-scales of detection on the same order but with a longer life-time allowing it to detect a larger volume of the sample which is not possible by the sensor alone.
 \end{abstract}
\maketitle
\section{Introduction}
Nuclear magnetic resonance (NMR) spectroscopy \cite{sc1, sc2} provides a label-free method for chemical analysis, which allows to gain detailed structural analysis as well as dynamical information from the specimen under study. Indeed, e.g. $80\%$ of all structure information on proteins stem from the method. However, NMR is rather insensitive and requires chemical purification of the structure under study. Much effort has been directed to applying NMR to nanoscale samples. Indeed, NMR detection of a $(4 nm)^3$ voxel of protons has been achieved with magnetic resonance force microscopy, a challenging experimental technique operating at ultralow temperature in vacuum \cite{sc3, sc4}. Under ambient conditions, microcoil detectors have enabled the detection of liquid samples of $(3000 {\rm nm})^3$ volume, but further miniaturization of this technique is not straightforward \cite{sc5, sc6}. Magnetic resonance imaging with few or single proton sensitivity would have significant impact on a number of research fields including life sciences and solid state physics. In addition if the detected spins do have good quantum properties they might become a valuable resource for quantum information protocols \cite{plenio}. 

Over the past decade single nitrogen-vacancy (NV) centers in diamond \cite{sc7,sc8, sc14, sc15} have been proposed and used as a novel atomic-size magnetic field sensor for detecting nanoscale nuclear spin ensembles or even single nuclear spins \cite{sc9}. This center is a joint defect in the carbon lattice of diamond, consisting of a substitutional nitrogen atom and an adjacent vacancy. Its spin triplet ($S=1$) groundstate can be polarized and read out optically, so
that electron spin resonance experiments can be performed on a single spin. A single center can be used as a nanoscale magnetic field sensor, able to detect a magnetic field in the nanotesla range in an integration time of $1s $ \cite{sc10}. This corresponds to the field of a single nuclear spin at a distance of a few nanometers.  Indeed, detection of single ${}^{13}C$ nuclear spins, and NMR signals from  a $(5-{\rm nm})^3$ voxel of various fluid and solid organic samples under ambient conditions has already been demonstrated \cite{dobro,sc12,sc13,sc,muller}. In these experiments the detection volume of the sample is shown to be limited by the life time of the sensor spin \cite{carlos1,carlos2}. In this paper we propose a detection scheme that is not limited by the $T_1$ of the sensor spin allowing for larger detection volumes on a given sample, and improving the sensitivity of resolving the distances between adjacent spins in the sample.
\begin{figure*}
\begin{center}
\includegraphics[width=0.9\textwidth]{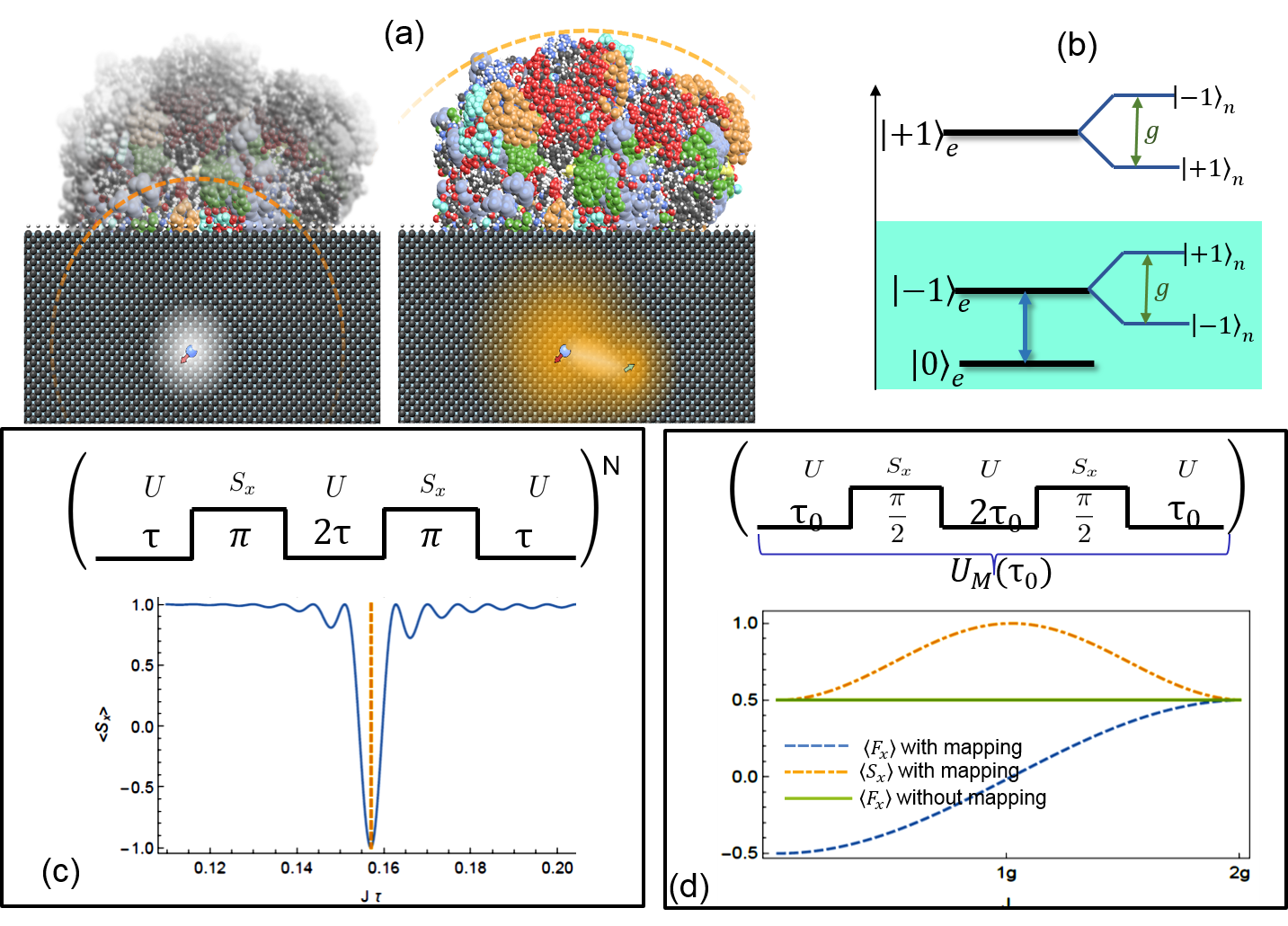}
\end{center}
\label{level}
\vspace{-5mm}
\caption{  \textbf{a} Schematic representation of the central idea, depicting the increase in detection volume of a sample on top of the diamond both through direct and indirect sensing. \textbf{b} The energy-levels of the sensor-ancillary system where only two levels of the sensor (within the triplet subspace) are resonantly coupled to a two-level ancillary (nuclear) spin. \textbf{c} We show the pulse sequence for a direct detection of a spin by the sensor. In the same figure we show the sharp transition of a sensor observable, $\langle S_x \rangle$ at $\tau = \pi/(2\omega_L + J)$. The parameters chosen for the simulation are $\omega_L/J =10$, $\theta = \pi/4$ and the optimal number of repetitions $N=23$. \textbf{d} Mapping pulse sequence $U_M(\tau_0)$ through which the sensor-sample coupling is directly reflected on the dynamics of the ancillary spin $F$. Also shown in the figure is the variation of the observables $\langle S_x \rangle$ and $\langle F_x \rangle$ both in the presence and absence of mapping.   }
\end{figure*}
In our prototype system, the NV center in diamond, the electronic spin interacts with two types of nuclear spins. One type are nuclear spins within the sample we want to measure. For those spins the electron spin acts as the quantum sensor sensing the fluctuations of the magnetic field caused by nearby spins in a given sample. Due to the distance dependent (dipolar) coupling between the sensor and these spins, the fluctuations of the later are reflected in observables changes of the sensor spin state. The other type of nuclear spins are those in the diamond material itself (e.g. ${}^{13}C$ or ${}^{14}N$) which are much stronger coupled to the NV spin, show long coherence times and thus can be used as ancilla quantum bits. While the strongly coupled spins, close to the sensor inside the diamond have dominant contribution to these changes, the weakly coupled sample ones (that are farther away) are mostly masked by the stronger ones. To resolve these weakly coupled spins in the sample, the strongly interacting spins typically are dynamically decoupled (DD) \cite{sc26, dd} so that the weak effects caused by farther spins are slowly accumulated in the sensor which is then finally readout \cite{sc13}. These weakly coupled sample spins can also become undetectable when their inverse coupling strength becomes comparable to the relaxation time of the sensor. While the masking of strongly interacting spins could be removed to a large extent by DD sequences, there is no way around to escape the relaxation time of the sensor placing an upper bound on the detectable spin volumes in the sample. To surpass this problem we propose an indirect or inductive method for sensing where a sensor spin is coupled to a long-lived ancillary spin (a strongly interacting nuclear spin) which now acts as a sensor for times longer than $T_1$ allowing one to detect larger parts of the sample. 

The physical setup for our proposed scheme consists of a sensor described by a three-level (spin-1) system that is interacting with the spins (that are typically spin-$1/2$ )of the sample. In the reference frame of the sensor one can think of the spins in the sample that are all distributed in a three-dimensional plane comprising of the $NV$ axis, (the direction along which the sensor is aligned) and the plane perpendicular to it. For simplicity we consider the two-dimensional configuration where the position of each spin in the sample is described by polar coordinates $r$ and $\theta$. We assume that the sensor is placed along the $z$-axis and is parallel to the applied magnetic field. In addition to the spins of the sample we also have an ancillary spin which is strongly interacting with the sensor along the $z$-axis. The relevant part of the Hamiltonian that determines the dynamics is simply given by 
\begin{equation}
H =\omega_L \sum_i I^z_i + gS^zF^z + S^z\sum_{i=1}^{N}J_i\left(\cos\theta_i I^z_i + \sin\theta_i I^x_i\right).
\end{equation}
where the sensor, ancillary and sample spins are respectively denoted by $S$, $F$ and $I$. The zero-field splitting of the sensor and the magnetic field strength $B_0$ can be chosen such that the sensor can act as a two-level system formed by the magnetic levels $m_S=0, ~ -1$ (see Fig. 1(b)). In the presence of field $B_0$ the sample spins precess at a Larmor frequency $\omega_L$.

\section{Indirect sensing by an ancillary spin}
In a standard detection scheme, using the sequence shown in Fig. 1c, \cite{dobro} the sensor is repetitively flipped (dynamically modulated) about the plane perpendicular to the $z$-axis at a rate that matches the energy of a particular sample spin $I_i$ which is $\omega_i = 2\omega_L + J_i \cos\theta_i$ \cite{dobro}. Under such matching conditions there is a transition in the sensor spin state which leads to a large observable effect in its optical or electrical readout (see Fig. 1c). The peak positions $\tau  = \pi/\omega_i$ depend on the parallel component ($z-z$) of the dipolar coupling between the sensor and the sample spins, while the width depends on the perpendicular component ($z-x$) of the interaction i.e., $\Delta \approx N/J_i\sin\theta_i$, where $N$ is the total number of pulse repetitions (see Fig. 1c). The optimal number of repetitions $N_{opt}$ required to obtain maximum contrast (height of the peak value) again depends on the values of $J_i$ and $\theta_i$. We will show later that in the presence of dissipation the peak width increases and the height decreases (thereby increasing the uncertainty of the detected position) but the peak position remains unchanged. Hence, by switching on a radio frequency (RF) field along $x$-direction one can exchange the roles of parallel and perpendicular components of dipolar coupling so that the peak now appears at $\tilde{\omega}_i = 2\omega_L + J_i \sin\theta_i$. Thus both the variables can be readout through the peak-positions only. 
This detection method can also be understood as the well known spectral overlap of the applied control (the filter function) on the sensor with the spectral distribution of the sample spins (Fig. 1c), a widely used technique in magnetic resonance imaging.

As can be seen, the above detection scheme can only resolve spins that are spectrally separated by distances larger than the width of the detection peak. Over the past years there have been quite some advances in improving this resolution by using more complex pulse sequences so as to have better resolution by avoiding cross-talk between the nearby spins, and also a high degree decoupling to the unwanted dephasing noise spectrum. But, the unavoidable relaxation of the sensor still remains a barrier, placing an upper bound on the detectable volumes of the sample. In the current proposal we surpass this effect by using a long-lived ancillary spin that is strongly coupled $g \gg J_i$ to the sensor. 

\begin{figure}
\includegraphics[width=90mm]{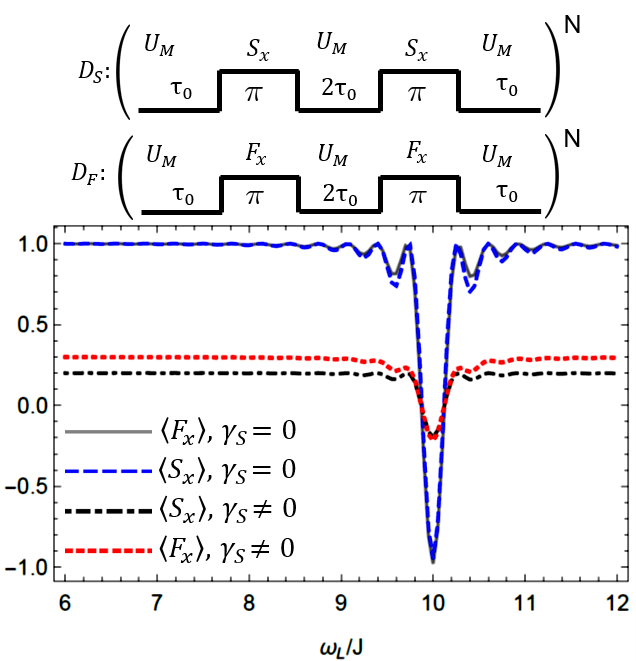}
\label{fds}
\vspace{-5mm}
\caption{Using the detection sequences $D_{S}$ and $D_F$ shown on the top of the figure we plot the expectation values of the sensor and ancillary spins $\langle S_x \rangle$ and $\langle F_x \rangle$ as a function of the Larmor frequency $\omega_L$ both in the presence and absence of relaxation. The sensor-ancillary coupling is much stronger than the sensor sample spin coupling and is taken to be $g = 80J$, and $\theta = \pi/4$. Relaxation is introduced through the Krauss form given in Eq. (5) with $\gamma_S/g=10^{-3}$. For consistency we have considered the relaxation of the sensor during the ancillary control pulses in the $D_F$ sequence as they are assumed to be considerably slower than the manipulations on the sensor spin. } 
\end{figure}

To make this ancillary spin which is not directly coupled to the sample spins also an indirect sensor, the first step would be to map the sensor-sample dynamics onto the ancillary spin. Though the below discussion is valid for any number of spins in the sample, for simplicity we shall concentrate on the case where the sample is described by a single spin coupled to the sensor with strength $J$. The dynamics of this three-spin system, the central spin (sensor) is coupled to the sample spin $I$ and to the ancillary spin $F$ is analytically solvable. Before we go ahead to design a mapping pulse sequence, we would like to first evaluate the behavior of the ancillary under free evolution generated by the Hamiltonian given in Eq. (1) as a function of $J$. One finds that the 
the time-dependent evolution of the coherence for the ancillary spin, given by
\begin{equation}
C_{F}(\tau) \equiv \langle F_x \rangle + i \langle F_y \rangle = \frac{1}{2}[1+\cos (g\tau)]C_F(0),
\end{equation}
has no dependence on the sample parameters $J$ and $\omega$ for any time $\tau$. This can be seen from the structure of the Hamiltonian (Eq. 1), where the ancillary-sensor coupling term commutes with the remainder and hence the sensor-sample coupling ($S-I$) only leads to a overall phase shift and does not contribute to the dynamics of the ancillary spin. The situation becomes different if a microwave field that couples energy states of the sensor is switched on. The total dynamics cannot be separated into the dynamics on $S-F$ and $S-I$ subspaces as these  terms no longer commute resulting in direct correlations between the ancillary and the sample. Instead of switching a continuous microwave field, we rather do it stroboscopically using the pulse sequence shown in Fig. 1(d). Under the action of this pulse sequence, the coherence of the ancillary spin now oscillates as a function of $J$ as shown in Fig. 1(d). For a fixed time $\tau_0 =\pi/g$ and $B_0 = 0$ the analytical expression for $C_F$ takes a simple form

\begin{equation}
C_{F}(\tau = \tau_0)  = \cos (J\tau_0) - i\cos(\theta)\cos(J\tau_0/2)\sin(J\tau_0).
\end{equation}
Now one can clearly see that on time-scales of $\tau_0$, the coherence of the ancillary spin purely oscillates on the frequency of the sample coupling parameters $J$ and $\theta$. For $J \ll g$ the purity of the ancillary state is close to one during the evolution, and changes as if it were directly coupled to the sample spin $I$. Equally the coherence of the sensor spin on this time scale only depends on $J$, given by
\begin{eqnarray}
C_{S}(\tau_0) &\equiv& \langle S_x \rangle + i \langle S_y \rangle \nonumber \\
&=&\frac{1}{2}(1+ \cos (J\tau_0)^2) - i\cos(\theta)\cos(J\tau_0/2)^2\sin(J\tau_0). \nonumber \\
\end{eqnarray}
\begin{figure}
\includegraphics[width=90mm]{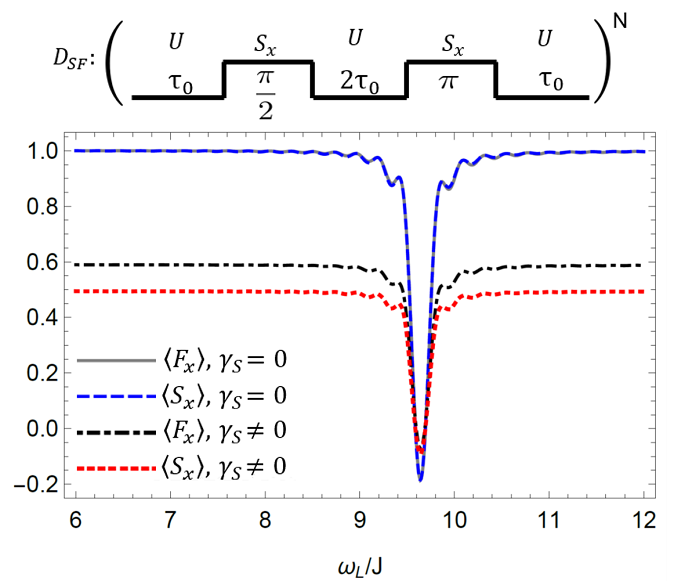}
\label{fds}
\vspace{-5mm}
\caption{Using the detection sequence $D_{SF}$ shown on the top of the figure we plot the expectation values of the sensor and ancillary spins $\langle S_x \rangle$ and $\langle F_x \rangle$ as a function of the Larmor frequency $\omega_L$ both in the presence and absence of relaxation. The sensor-ancillary coupling is much stronger and is taken to be $g = 80J$, and $\theta = \pi/4$. Relaxation is introduced through the Krauss form given in Eq. (5) with ${\rm e}^{-\gamma_S\tau_0} = 10^{-3}$. } 
\end{figure}
We would like to note that by using the above mapping sequence the time units are now renormalized in multiples of $\tau_0$. For example, with a Larmor frequency $\omega_L$ and interaction $J$, a detection sequence shown in Fig. 1(c) would give a sharp transition in the sensor coherence at $\tau =  \pi/2\omega_L + J \cos\theta $. But with the mapping sequence, $\tau$  should now be varied in steps of $\tau_0$, implying that spins whose total Zeeman energy is smaller than the $S-F$ coupling strength $g$ are detectable and the parameters should satisfy $g > \omega_L > J$, a commonly available situation in real physical systems. 
Having mapped the parameters of the sample onto the ancillary spin, the next step is to detect the sample spin by the ancillary. One should also remember that the sample spin could also be detected by the sensor itself as its coherence is also a function of $J$ and $\theta$ during every mapping sequence. One can see from Fig. 2 that we now have the possibility to sense the sample spin either by running the detection sequence on the sensor $D_S$ or a detection sequence on the ancillary spin $D_F$, and see an almost identical transitions in their observables $\langle S_x \rangle$ and $\langle F_x \rangle$ respectively. Though the difference is quite small one can find that sensing on the ancillary has a better contrast than on the sensor, and this amplifies in the presence of relaxation which we shall discuss later. The advantage of the mapping sequence becomes more evident when the sensor needs to be reinitialized quite frequently, making it ineligible to sense the weak sample spin, but the ancillary spin can still reflect the same transition independent of sensor's reinitialization rate in the absence of any relaxation.
If we now squeeze both the mapping and detection into one sequence, allowing for control only on the sensor (as they could be quite faster than controlling the ancillary) but still achieving the goal of reflecting the information of the sample spin on both the sensor and ancillary could be quite useful. For this we use a single asymmetric sequence (shown in Fig. 3) where a $\pi/2$-pulse (for mapping) and $\pi$ (for detecting) are performed on the sensor during the first and second half of the sequence. Surprisingly, repeating this single sequence leads to identical dynamics on both the sensor and ancillary in the absence of relaxation. This can be seen from Fig. 3 where we have plotted  the observables $\langle S_x \rangle$ and $\langle F_x \rangle$ that overlap perfectly. 

We now turn to a realistic situation where the sensor is relaxing to its thermal equilibrium at a rate $\gamma_S$. Assuming a pure Markovian decay, we introduce relaxation as a depolarizing channel \cite{nc} through the Krauss form such that the sensor spin state at any time $t$ is simply given by
\begin{equation}
\mathcal{E}(\rho(t)) = \frac{1}{2}\left(1-{\rm e}^{-\gamma_St}\right) \mathbbm{1}\otimes{\rm Tr}_S(\rho(t)) + {\rm e}^{-\gamma_St}\rho(t),
\end{equation}
where $\rho(t)$ is the density matrix of the total system at any time $t$.
Though the above way of introducing relaxation into the problem is simpler than solving the master equation for the total system exactly, it still could capture most of the physics relevant to the current discussion.

\begin{figure}
\hspace{-5mm}
\includegraphics[width=90mm]{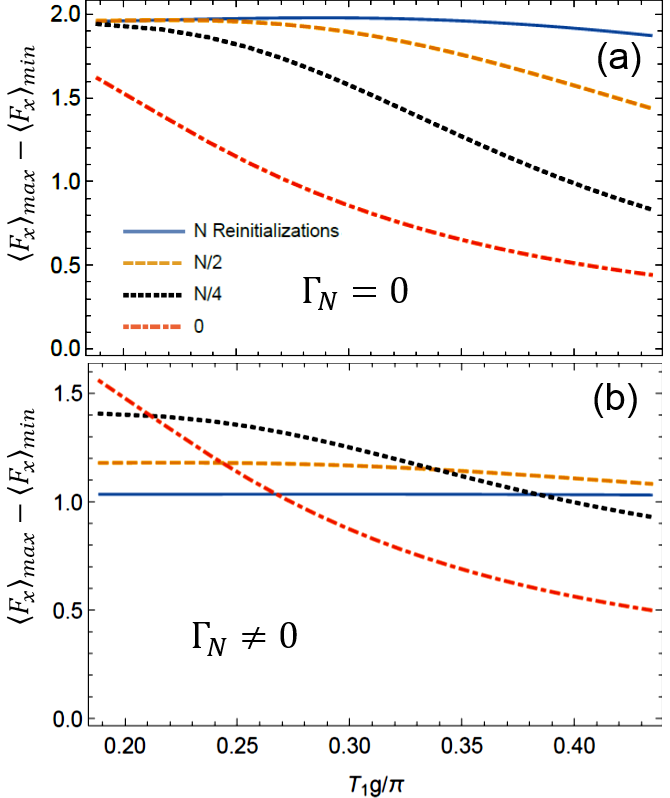}
\label{fds}
\vspace{-2mm}
\caption{The maximum contrast of the ancillary spin observable is plotted as function of the relaxation time for different reinitialization rates for two cases (a) when the sensor reinitialization does not dephase the ancillary spin and (b) when the sensor reinitialization dephases the ancillary spin at a rate $\Gamma_N$.  The dephasing $\Gamma_N$ is chosen such that ${\rm e}^{-\Gamma_N \tau_0} = 10^{-2}$, where the sensor-ancillary coupling and other parameters are similar to that given in earlier plots. } 
\end{figure}

From the above analysis we have already seen that indirect sensing through the ancillary spin shows better contrast in the presence of relaxation even though they have equal contrast in the absence of relaxation. We now want to know if the detection through ancillary spin could be made independent of the sensor relaxation $\gamma_S$. We have already seen from Fig. 2 that using the mapping sequence, an independent detection of the sample spin can be made through the ancillary using $D_F$. While running the detection sequence on the ancillary even if the sensor is reinitialized though it destroys the sensor-ancillary ($S-F$) correlations, the ancillary-sample ($F-I$) remain intact. Hence the availability of a fully coherent sensor spin state at the end of each repetition would map the sample spin parameters onto the ancillary with almost unit efficiency (as in the ideal case of $\gamma_S=0$). Instead if the reinitializations are performed at a slower rate, the relaxation effects of the sensor enter the dynamics of the ancillary, thereby reducing its contrast from the ideal value. This is shown in Fig. 4(a), where we plot the contrast observed on the ancillary spin readout as a function of the relaxation time $T_1 = 1/\gamma_S$, for different initializations of the sensor. Under very frequent initializations the contrast remains independent of $T_1$, but for slower rates the loss of purity occurred till the reinitialization decreases the overall contrast.
In a realistic situation one should also consider an additional dephasing incurred by the ancillary during the reinitialization of the sensor. This happens for example due to the difference in the hyperfine coupling between the sensor and the ancillary spin in the ground and excited states. Taking into account this additional dephasing, one immediately finds that frequent reinitialization may not be the ideal situation, which can be seen from Fig. 4(b).

We now make some numerical estimates on the new range of distant spins that the ancillary spin could sense. For NV centers in diamond one can find an ancillary spin such that the sensor-ancillary coupling is $g\approx1$MHz, and the sensor has a typical relaxation rate $\gamma_S=1$kHz. The sensitivity with which a sample spin whose coupling strength $J =10$kHz could be detected reduces to $17\%$ of its maximum achievable case when $\gamma_S=0$. Now running the detection sequence on the ancillary spin, with the possibility to make a $\pi$ flip at rate $g$, we find that the sample spin could now be detected with an increased sensitivity of  $25\%$. Now if we reinitialize the senor spin at a rate of $g/10$ i.e., the sensor is reinitialized three times in the entire sequence at very $10 \mu$s, increases the sensitivity to $~33\%$. The dephasing incurred by the ancillary (nuclear) spin during the reinitialization process is proportional to the reinitialization time of the sensor (electron) spin $\tau_r$ and the hyperfine coupling between the ancillary and the sensor in the excited state $g_e$. While choosing strongly interacting nuclei as ancillary spin will improve the spectral resolution of the sample spins, it equally will get dephased fast enough during frequent reinitialization of the sensor and vice versa. Thus finding an optimally coupled ancillary spin would become a key requirement for the indirect sensing proposed in this paper.

\section{Conclusion}In conclusion we have proposed a new method to detect distant spins that are very weakly coupled to the sensor. When this coupling becomes much smaller than the relaxation rate of the sensor, direct sensing yields almost no contrast, while an indirect sensing through a long-lived ancillary spin coupled strongly to the sensor can be used to detect the same spins with a high contrast. We have shown that this kind of dual sensing could be achieved by a mapping sequence between the sensor and the ancillary so that the information related to the sample spin is stored in both the spins, and becomes detectable by either of them. Though the analysis is made on a simple model system having a single sample spin, the analysis could be straightforwardly generalized to highly dense samples with few hundreds of the sample spins. 

\section*{Acknowledgements}
We thank Ingmar Jakobi for providing illustrative graphics and Sebastian Zaiser, Amit Finkler for useful comments and discussions. We acknowledge the financial support by the ERC project SQUTEC, DFG (FOR1693), DFG SFB/TR21, EU DIADEMS, SIQS, Max Planck Society and the Volkswagenstiftung. 

\bibliography{weaklycoupled} 

\end{document}